\documentstyle[twocolumn,epsf]{jpsj}
\def\runtitle{Periodic Anderson model
with degenerate orbitals}
\def\runauthor{Ryota   {\sc Sato}, Takuma {\sc Ohashi}, Akihisa {\sc Koga} 
and Norio {\sc Kawakami}
} 
\title{ Periodic Anderson model with degenerate orbitals:\\
linearized dynamical mean field theory approach}
\author{Ryota {\sc  Sato},
Takuma {\sc Ohashi},
Akihisa {\sc Koga},
and Norio {\sc Kawakami}
}
\inst{Department of Applied Physics,
Osaka University, Suita, Osaka 565-0871, Japan}

\recdate{\today}

\abst{%
We investigate a multi-orbital extension of the 
 periodic Anderson model with particular emphasis 
on electron correlations including orbital fluctuations.
By means of a linearized version of the dynamical mean-field theory, 
we compute the renormalization 
factor,  the density of states,  the spectral gap and 
the local correlation functions for
 a given set of the intra- and inter-orbital Coulomb interactions 
as well as the Hund coupling. 
It is found that when a certain condition is met for the 
intra- and inter-orbital interactions for $f$ electrons,
 orbital fluctuations are
enhanced, thereby enlarging the Kondo insulating gap.
This effect is suppressed in the presence of the 
Hund coupling.  We also clarify how the Kondo insulator
is continuously changed to the Mott insulator when 
electron correlations among conduction electrons are 
increased.
}

\kword
{heavy fermions, periodic Anderson model, orbital degeneracy,
dynamical mean field theory, 
}

\begin{document}
\sloppy
\maketitle

%%%%%%%%%%%%%%%%%%%%%%%%%%%%%%%%%%%%%%%%%%%%%%%%%%%%%%%%%%%%%%%%%%%
%%%%%%%%%%%%%%%%%%%%%%%%%%%%%%%%%%%%%%%%%%%%%%%%%%%%%%%%%%%%%%%%%%%
%%%%%%%%%                     %%%%%%%%%%%%%%%%%%%%%%%%%%%%%
%%%%%%%%% 1.  Introduction    %%%%%%%%%%%%%%%%%%%%%%%%%%%%%
%%%%%%%%%                     %%%%%%%%%%%%%%%%%%%%%%%%%%%%%
%%%%%%%%%%%%%%%%%%%%%%%%%%%%%%%%%%%%%%%%%%%%%%%%%%%%%%%%%%%%%%%%%%%
%%%%%%%%%%%%%%%%%%%%%%%%%%%%%%%%%%%%%%%%%%%%%%%%%%%%%%%%%%%%%%%%%%%

\section{Introduction}
Since the discovery of heavy-fermion materials with rare-earth
 or actinide elements, comprehensive understanding of 
 this class of correlated electron 
systems has received considerable attention. 
In these compounds, strong correlations among 
$f$ electrons play an important role to
 form the heavy quasi-particle state at low temperatures.
In particular, if correlated $f$-electron systems are
in the insulating phase, they are referred to as the Kondo
insulator, which possesses a wide variety of compounds.
\cite{T.Susaki,K.Sugiyama,T.Saso-1,M.F.Hundley,A.Severing,
P.S.Riseborough,C.S.Castro,
J.W.Allen,J.C.Cooley,T.E.Mason}
To discuss the electronic properties
theoretically,  the periodic Anderson model (PAM)
has been investigated extensively. This model
 is simplified to extract the essence of 
heavy fermions, which is usually  described by 
free conduction electrons coupled to
highly correlated single-orbital $f$ electrons.

In some heavy-fermion compounds, the
other parameters dropped in the original PAM,
 such as the orbital degeneracy and the interactions among 
 conduction electrons,
are also important to explain the experimental findings.
For example, the compound Nd$_{2-x}$Ce$_x$CuO$_4$
\cite{T.Brugger,J.Igarashi,Y.M.Li,T.Schork,K.Itai} 
shows the unusual behavior in the specific heat, 
which may be explained by properly taking into account 
the correlations due to not only
  $f$ electrons but also conduction electrons.
%For instance, it is referred that 
%the effect of correlated conduction electrons have to take into 
%consideration for heavy-fermion behavior found in Nd$_{2-x}$Ce$_x$CuO$_4$, 
%\cite{T.Brugger,J.Igarashi,Y.M.Li,T.Schork,K.Itai} 
%and for the formation mechanism of hybridization gap 
%in Kondo insulator YbBe$_{12}$, \cite{T.Susaki}
 Furthermore, it is claimed that the orbital degrees of freedom affect
 the Kondo-insulating gap around Fermi surface
for another prototype of heavy fermion 
compound,  $\rm YbBe_{12}$.\cite{T.Susaki,K.Sugiyama,T.Saso-1}
These findings naturally encourage us to explore the effects 
beyond the simple PAM systematically,  such as correlations due 
to conduction electrons, the orbital effects,
etc.

In this paper, we investigate 
the doubly degenerate periodic Anderson model
to discuss the effects on the Kondo insulator 
due to orbital degeneracy together with 
electron correlations for $f$ and
conduction bands.  By exploiting a linearized version of
dynamical mean field theory (DMFT),
we  show that the interplay of several interactions 
yields interesting effects on the formation of the Kondo insulator,
which do not appear in the simple PAM.

%In the DMFT, lattice model is mapped to an effective impurity model. 
%The DMFT, which becomes exact in infinite dimensions, requires the 
%self-consistent solution of a single-impurity Anderson model(SIAM). 
%Recently the new approach to solve the SIAM is suggested by Potthoff. 
%This approach simplifies the SIAM, however keeps the essential feature 
%of the DMFT. 
%We have extended the simplified dynamical 
%mean-field theory to this model, and treated in this framework. 

This paper is organized as follows. 
In \S 2, we introduce the model Hamiltonian
with orbital degeneracy, and 
 then briefly explain a linearized version of DMFT.
In \S 3, we  discuss electron correlations due to
conduction bands  by employing 
the single-orbital model. We then explore in \S 4 
how the interplay of several distinct interactions 
in the degenerate  model
produces nontrivial effects on the low-energy electronic 
properties.
A brief summary is given in the last section.

%%%%%%%%%%%%%%%%%%%%%%%%%%%%%%%%%%%%%%%%%%%%%%%%%%%%%%%%%%%%%%%%%%%
%%%%%%%%%%%%%%%%%%%%%%%%%%%%%%%%%%%%%%%%%%%%%%%%%%%%%%%%%%%%%%%%%%%
%%%%%%%%%                                %%%%%%%%%%%%%%%%%%%%%%%%%%
%%%%%%%%% 2. Model and Formulation       %%%%%%%%%%%%%%%%%%%%%%%%%%
%%%%%%%%%                                %%%%%%%%%%%%%%%%%%%%%%%%%%
%%%%%%%%%%%%%%%%%%%%%%%%%%%%%%%%%%%%%%%%%%%%%%%%%%%%%%%%%%%%%%%%%%%
%%%%%%%%%%%%%%%%%%%%%%%%%%%%%%%%%%%%%%%%%%%%%%%%%%%%%%%%%%%%%%%%%%%
\section{ Model Hamiltonian and Method }

\subsection{ Two-orbital periodic Anderson model}

We consider the periodic Anderson model with 
two-fold degenerate orbitals, 
which may be described by the following Hamiltonian, 
%%%%%%%%%%%%%%%%%%%%%%%%%%%%%%%%%%%%%%%%%%%%%%%%%%%%%%%%%%%%%%%%%%%%%%%%%%%%
\begin{eqnarray}
{\cal H}_{PAM}&=&{\cal H}_{0}+{\cal H}_{I} \label{lattice} \\
{\cal H}_{0}
&=& %\sum_{m=1}^{2} \Bigg [
    \sum_{\stackrel{<i,j>}{m,\sigma}} t_{ij}c^{\dag}_{im\sigma}c_{jm\sigma}
  + \sum_{i,m,\sigma} \epsilon_{f}f^{\dag}_{im\sigma}f_{im\sigma} \nonumber \\
&+& V \sum_{i,m,\sigma}\left( f^{\dag}_{im\sigma}c_{im\sigma}+h.c.\right) 
%\Bigg ] 
\\
{\cal H}_{I}
&=& U_{f}\sum_{i,m}n^{f}_{im\uparrow}n^{f}_{im\downarrow}
 +  U_{c}\sum_{i,m}n^{c}_{im\uparrow}n^{c}_{im\downarrow}
\nonumber \\
&+& U^{\prime}_{f}\sum_{i,\alpha,\beta}
    n^{f}_{i1\alpha}n^{f}_{i2\beta}
 +  U^{\prime}_{c}\sum_{i,\alpha,\beta}
    n^{c}_{i1\alpha}n^{c}_{i2\beta}
\nonumber \\
&-& J_{f}\sum_{i}{\bf S}^{f}_{i1} \cdot {\bf S}^{f}_{i2}
 -  J_{c}\sum_{i}{\bf S}^{c}_{i1} \cdot {\bf S}^{c}_{i2}
\label{Hloc}
\end{eqnarray}
%%%%%%%%%%%%%%%%%%%%%%%%%%%%%%%%%%%%%%%%%%%%%%%%%%%%%%%%%%%%%%%%%%%%%%%%%%%%%
where $f_{im\sigma}  (c_{im\sigma})$ annihilates an $f$ electron 
 (conduction electron) with spin $\sigma(=\uparrow,\downarrow)$ 
and orbital $m(=1,2)$ at the $i$th site, 
$n_{im\sigma}^{c}=c_{im\sigma}^\dag c_{im\sigma}$ and 
$n_{im\sigma}^{f}=f_{im\sigma}^\dag f_{im\sigma}$.
Here, $t_{ij}$ represents the hopping integral, 
$\epsilon_f$ the energy level of the 
$f$ state, and $V$ the hybridization between the conduction and 
$f$ states.
$U_{f} (U_c)$ is the intra-orbital Coulomb interaction, 
$U^{\prime}_{f} (U^{\prime}_{c})$ is the inter-orbital Coulomb interaction, 
and $J_{f} (J_{c})$ is Hund coupling for
$f$ electrons  (conduction electrons).
The spin operators are 
defined by ${\bf S}_{im}^f=\frac{1}{2} \sum_{\alpha \beta} 
f^{\dag}_{im\alpha} {\bf \tau}_{\alpha \beta} f_{im\beta}$ 
and 
${\bf S}_{im}^c=\frac{1}{2} \sum_{\alpha \beta} 
c^{\dag}_{im\alpha} {\bf \tau}_{\alpha \beta} c_{im\beta}$,
where ${\bf \tau}$ is the Pauli matrix. 
Note that the model has two orbitals 
 both for conduction and $f$ electrons, which are specified 
by the same index, $m$. This scheme has been
sometimes used for investigating the orbital effects
on heavy fermion systems, which may capture some 
essential properties due to degenerate orbitals.
\cite{Y.Ono-1,A.Tsuruta,M.Jarrell-1}
In this paper,  we 
 discuss how electron correlations are developed 
in the Kondo insulator with multi-orbitals, by changing 
 the parameters 
$\{ U_f, U_f', J_f\}$ and $\{ U_c, U_c', J_c\}$ systematically.

%%%%%%%%%%%%%%%%%%%%%%%%%%%%%%%%%%%%%%%%%%%%%%%%%%%%%%%%%%%%%%%%%%%
%%%%%%%%%                         %%%%%%%%%%%%%%%%%%%%%%%%%
%%%%%%%%% 2. Two-site DMFT         %%%%%%%%%%%%%%%%%%%%%%%%%
%%%%%%%%%                         %%%%%%%%%%%%%%%%%%%%%%%%%
%%%%%%%%%                        %%%%%%%%%%%%%%%%%%%%%%%%%
%%%%%%%%%%%%%%%%%%%%%%%%%%%%%%%%%%%%%%%%%%%%%%%%%%%%%%%%%%%%%%%%%%%
%%%%%%%%%%%%%%%%%%%%%%%%%%%%%%%%%%%%%%%%%%%%%%%%%%%%%%%%%%%%%%%%%%%

\subsection{Linearized dynamical mean-field theory}

We make use of DMFT, which has
been developed by several groups\cite{W.Metzner,MullerHartmann,Rev,Pruschke} 
and 
has successfully been applied 
to the single-band Hubbard model,
\cite{W.Metzner,M.Caffarel,X.Y.Zhang,A.Georges,M.Jarrell-2,M.Potthoff-1,R.Bulla-1,R.Bulla-2,Y.Ono-2,H.Kajueter,D.S.Fisher}
the two-band Hubbard model,
\cite{M.Caffarel,W.Metzner2,Y.Imai,A.Koga1,A.Koga2,Y.Ono,T.Momoi,M.J.Rozenberg,K.Held,J.E.Han,V.S.Oudovenko,Liebsch,A.Koga3,S.Florens}
the PAM \cite{M.Jarrell-3,Th.Pruschke,T.Mutou,
T.Saso-2,D.Meyer,T.Schork} , etc. 
This treatment can take into account local electron correlations precisely, 
and thereby is exact in infinite dimensions. Then
the lattice model can be mapped onto an effective impurity model,
which has been solved by a variety of methods such as 
the iterated perturbation theory, \cite{X.Y.Zhang,H.Kajueter,Liebsch}
the non-crossing approximation, \cite{C.I.Kim,Y.Imai,T.Schork}
the projective self-consistent method. \cite{D.S.Fisher,S.Florens}
Since these methods are not efficient enough to treat the systems 
with degenerate orbitals, powerful
numerical methods have been employed, {\it e.g.} 
the exact diagonalization, \cite{T.Momoi,Y.Ono,A.Koga1,A.Koga3}
the quantum Monte Carlo simulation.
\cite{M.J.Rozenberg,K.Held,J.E.Han,V.S.Oudovenko,A.Koga2,Liebsch}
Recently a new method to solve the 
effective impurity model has been proposed by Potthoff,\cite{M.Potthoff-1}
which we will refer to the linearized DMFT in this paper.
This approach simplifies the procedure of DMFT by linearizing 
the self-consistent equations in the low-energy region,
but still keeps the essential features
of electron correlations. In this 
approximation, the effective bath is represented by
 a few sites. In spite of this simplification, 
electronic properties in the low-energy region 
around the Fermi surface can be described rather well.\cite{R.Bulla-2,Y.Ono-2}
In fact, the critical values of the Hubbard model with and without 
degenerate orbitals are in good agreement with the other numerical 
techniques.\cite{R.Bulla-1,A.Koga1}

In this paper, we exploit the 
linearized DMFT, which  may convert
 the original model eq. (\ref{lattice})
to the effective impurity Anderson model,
%%%%%%%%%%%%%%%%%%%%%%%%%%%%%%%%%%%%%%%%%%%%%%%%%%%%%%%%%%%%%%%%%%%%%%%%%%
\begin{eqnarray}
{\cal H}_{imp}&=& 
\sum_{lm\sigma}\tilde{\varepsilon}_{l}d^{\dag}_{lm\sigma}d_{lm\sigma}
+ \sum_{lm\sigma}\tilde{V}_l \left( d^{\dag}_{lm\sigma}c_{m\sigma}+h.c. \right)
\nonumber \\
&+& \tilde{\varepsilon}_{c}\sum_{lm\sigma}c^{\dag}_{m\sigma}c_{m\sigma}
 +  \tilde{\varepsilon}_{f}\sum_{lm\sigma}f^{\dag}_{m\sigma}f_{m\sigma}
\nonumber\\
&+& V\sum_{lm\sigma}\left( f^{\dag}_{m\sigma}c_{m\sigma}+h.c. \right)
\nonumber \\
&+& U_{f}\sum^{2}_{m=1}n^{f}_{m\uparrow}n^{f}_{m\downarrow}
 +  U_{c}\sum^{2}_{m=1}n^{c}_{m\uparrow}n^{c}_{m\downarrow}
\nonumber \\
&+& U^{\prime}_{f}\sum_{\alpha,\beta}
    n^{f}_{1\alpha}n^{f}_{2\beta}
 +  U^{\prime}_{c}\sum_{\alpha,\beta} 
    n^{c}_{1\alpha}n^{c}_{2\beta}
\nonumber \\
&-& J_{f}{\bf S}^{f}_{1} \cdot {\bf S}^{f}_{2}
 -  J_{c}{\bf S}^{c}_{1} \cdot {\bf S}^{c}_{2}
\label{s-imp}
\end{eqnarray}
%%%%%%%%%%%%%%%%%%%%%%%%%%%%%%%%%%%%%%%%%%%%%%%%%%%%%%%%%%%%%%%%%%%%%%%%%%
where $\tilde{\epsilon}_c$ and $\tilde{\epsilon}_f$ are the
renormalized energy levels,
 which should be determined by the number of
conduction and $f$ electrons in the system. Here,
$d^{\dag}_{lm\sigma} (d_{lm\sigma})$ creates (annihilates) an electron 
in the effective bath with two sites $(l=1,2)$. 

To determine the effective energy
$\tilde{\varepsilon}_l$ and the hybridization $\tilde{V}_l$,
we linearize the self-energies in the small $\omega$ region as,
%%%%%%%%%%%%%%%%%%%%%%%%%%%%%%%%%%%%%%%%%%%%%%%%%%%%%%%%%%%%%%%%%%%%%%%%%%
\begin{equation}
\begin{array}{rcl}
\Sigma^{ff}(\omega) &=& a+b\omega+O(\omega^{2}) \\
\Sigma^{cc}(\omega) &=& c+d\omega+O(\omega^{2}) \\
\Sigma^{fc}(\omega) &=& g+h\omega+O(\omega^{2}),
%\Sigma^{f_{1}f_{2}}(\omega) &=& i+j\omega+O(\omega^{2})
\end{array}
\label{self-energy}
\end{equation}
%%%%%%%%%%%%%%%%%%%%%%%%%%%%%%%%%%%%%%%%%%%%%%%%%%%%%%%%%%%%%%%%%%%%%%%%%%
where $a, b, c, d, g$ and $h$ are real numbers to be determined.
The renormalization factors for $f$ and
conduction electrons  are respectively given as,
%%%%%%%%%%%%%%%%%%%%%%%%%%%%%%%%%%%%%%%%%%%%%%%%%%%%%%%%%%%%%%%%%%%%%%%%%%
\begin{eqnarray}
Z_{f} &=& (1-b)^{-1} \\
Z_{c} &=& (1-d)^{-1}. 
%%Z_{fc} &=& (1-h)^{-1}
\label{reno}
\end{eqnarray}

On the other hand, the Green function for the lattice system is given as,
%%%%%%%%%%%%%%%%%%%%%%%%%%%%%%%%%%%%%%%%%%%%%%%%%%%%%%%%%%%%%%%%%%%%%%%%%%
\begin{eqnarray}
G(z)&=&\int dk G(k,z)\\
G^{-1}(k, z)
&=&\left(
\begin{array}{cc}
z-\varepsilon_{k}-\Sigma^{cc}(z) & -V-\Sigma^{cf}(z) \\
-V-\Sigma^{fc}(z) & z-\varepsilon_{f}-\Sigma^{ff}(z) \\
\end{array} \right). \nonumber\\
\label{G}
\end{eqnarray}
%%%%%%%%%%%%%%%%%%%%%%%%%%%%%%%%%%%%%%%%%%%%%%%%%%%%%%%%%
We consider here the $D-$dimensional Bethe lattice 
with the hopping integral $t=t^{\ast}/\sqrt{D}$, which results in 
the density of states,
%%%%%%%%%%%%%%%%%%%%%%%%%%%%%%%%%%%%%%%%%%%%%%%%%%%%%%%%%%%%%%%%%%%%%%%%%%%%%
\begin{eqnarray}
\rho_{0}(z)=\frac{1}{2\pi {t^{\ast}}^{2}}\sqrt{4{t^{\ast}}^{2}-z^{2}}. 
\end{eqnarray}
%%%%%%%%%%%%%%%%%%%%%%%%%%%%%%%%%%%%%%%%%%%%%%%%%%%%%%%%%%%%%%%%%%%%%%%%%%%%%
By comparing the local Green function with the impurity Green function, 
the self-consistent equation for DMFT now reads
%%%%%%%%%%%%%%%%%%%%%%%%%%%%%%%%%%%%%%%%%%%%%%%%%%%%%%%%%%%%%%%%%%%%%%%%%
\begin{eqnarray}
\left[ G^{-1}_{0,imp}(z) \right]_{cc}
 = z-\left(\frac{W}{4} \right)^{2}G^{cc}_{loc}(z),
\label{self}
\end{eqnarray}
where $W(=4t^{\ast})$ is the band width.

By substituting the equation (\ref{self-energy}) 
to the self-consistent equations, 
we end up with  the hybridization and the energy of  $f$ level for
the effective bath, 
%%%%%%%%%%%%%%%%%%%%%%%%%%%%%%%%%%%%%%%%%%%%%%%%%%
\begin{eqnarray}
\tilde{V}_l^2&=&\left(\frac{W}{4}\right)^2 \frac{Z_{c}}{2}\label{new-self-1}\\
\tilde{\varepsilon}_{l}^2&=&V_{g}^{2}+M^{(0)}_{2}{Z_{c}}^{2},
\label{new-self-2}
\end{eqnarray}
%%%%%%%%%%%%%%%%%%%%%%%%%%%%%%%%%%%%%%%%%%%%%%%%%%%
where $M^{(0)}_{2}=\Sigma_{j\neq i}t^{2}_{ij}=\int dx \ x^{2}\rho_{0}(x)$ 
is the variance of the noninteracting density of states 
and $V_g \equiv V+g$ is the effective hybridization.
%We assume here $h = 0$, since $h \sim 0$ always holds 
%in our parameter regime. 
We note here that the quantities $a$ and $c$
in (\ref{self-energy}) can be incorporated
in $\tilde{\epsilon}_f$ and $\tilde{\epsilon}_c$,
and assume $h=0$ since $h\sim 0$ always holds
in the parameter regime we are now interested in.

In the following, we take  $t^*=1$ as unit of the energy and 
fix the hybridization $V=1.0$. 
We deal with the symmetric case in the PAM (half-filled bands) 
by setting $\tilde{V}_1=\tilde{V}_2=\frac{W}{4}\sqrt{\frac{Z_{c}}{2}}$, 
$\tilde{\varepsilon}_1=-\tilde{\varepsilon}_2=
\sqrt{V_{g}^{2}+M^{(0)}_{2}{Z_{c}}^{2}}$, $\tilde{\varepsilon}_c=-U_c/2-U'_c$,
and $\tilde{\varepsilon}_f=-U_f/2-U'_f$, for simplicity.

%%%%%%%%%%%%%%%%%%%%%%%%%%%%%%%%%%%%%%%%%%%%%%%%%%%%%%%%%%%%%%%%%%%
%%%%%%%%%%%%%%%%%%%%%%%%%%%%%%%%%%%%%%%%%%%%%%%%%%%%%%%%%%%%%%%%%%%
%%%%%%%%%                        %%%%%%%%%%%%%%%%%%%%%%%%%%
%%%%%%%%% 3. results                     %%%%%%%%%%%%%%%%%%%%%%%%%%
%%%%%%%%%                        %%%%%%%%%%%%%%%%%%%%%%%%%%
%%%%%%%%%                       %%%%%%%%%%%%%%%%%%%%%%%%%%
%%%%%%%%%%%%%%%%%%%%%%%%%%%%%%%%%%%%%%%%%%%%%%%%%%%%%%%%%%%%%%%%%%%
%%%%%%%%%%%%%%%%%%%%%%%%%%%%%%%%%%%%%%%%%%%%%%%%%%%%%%%%%%%%%%%%%%%

%%%%%%%%%%%%%%%%%%%%%%%%%%%%%%%%%%%%%%%%%%%%%%%%%%
\section{Single-Orbital Periodic Anderson Model}
%%%%%%%%%%%%%%%%%%%%%%%%%%%%%%%%%%%%%%%%%%%%%%%%%%

We begin with the single-orbital PAM.
As is well known, in the conventional PAM 
with $U_c=0$, 
the increase of $U_f$ 
enhances correlations among $f$ electrons, 
resulting in the heavy quasi-particle state.
For half-filled bands considered here, 
the system naturally leads to the Kondo insulator with 
a renormalized spectral gap.
Here, we focus on the role played by
 electron correlations due to the conduction band.
To this end, 
we calculate  the renormalization factors
 $Z_c$ and $Z_f$ as 
a function of the {\it c-c} Coulomb interaction $U_c$.
%%%%%%%%%%%%%%%%%%%%%%%%%%%%%%%%%%%%%%%%%%%%%%%%%%%%%%%%%%%%%%%%%%%
\begin{figure}[htb]
\begin{center}
\vskip 5mm
\leavevmode \epsfxsize=80mm 
\epsffile{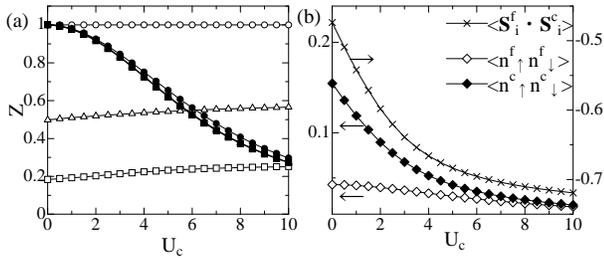}
\end{center}
\caption{(a) Renormalization factors as a function of 
the {\it c-c} Coulomb interaction $U_c$:
solid (open) symbols represent $Z_{c}$ ($Z_f$). 
We show the results for $U_f=0.0, 5.0$ and $10.0$ as 
circles, triangles and squares, respectively. 
(b) Local spin correlation function 
$\langle {\bf S}_i^f\cdot {\bf S}^c_i \rangle$, and
double-occupation probabilities 
$\langle n^f_\uparrow n^f_\downarrow \rangle$ 
and $\langle n^c_\uparrow n^c_\downarrow \rangle$.
}
\label{fig:single-renor}
\end{figure}
%%%%%%%%%%%%%%%%%%%%%%%%%%%%%%%%%%%%%%%%%%%%%%%%%%%%%%%%%%%%%%%%%%%
As shown in Fig. \ref{fig:single-renor} (a), the introduction 
of $U_c$ reduces  the renormalization factor $Z_c$, while
it does not alter the factor $Z_f$ so much.
Namely, the {\it c-c} interaction $U_c$ mainly renormalizes
 conduction electrons, as naively expected.
We also show in Fig. \ref{fig:single-renor} (b) 
 the {\it c-f} spin correlation function, 
$\langle {\bf S}^{f}_{i}\cdot {\bf S}^{c}_{i}\rangle$, 
and the double-occupation probabilities,
$\langle n^f_\uparrow n^f_\downarrow \rangle$ 
and $\langle n^c_\uparrow n^c_\downarrow \rangle$,
for  $f$ and conduction electrons.
%%%%%%%%%%%%%%%%%%%%%%%%%%%%%%%%%%%%%%%%%%%%%%%%%%%%%%%%%%%%%%%%%%%%%%%%%%%%%%%
%\begin{eqnarray}
%\langle n^{f}_{m\uparrow}n^{f}_{m\downarrow} \rangle
%&=&< n^{f}_{\uparrow}n^{f}_{\downarrow} > \\
%\langle n^{c}_{m\uparrow}n^{c}_{m\downarrow} \rangle
%&=&< n^{c}_{\uparrow}n^{c}_{\downarrow} >
%\left< {\bf S}^{f}_{Z}\cdot {\bf S}^{c}_{Z} \right>&=& -\frac{3}{4}
%\left< \left( n^{f}_{\uparrow}-n^{f}_{\downarrow} \right)
%                \left( n^{c}_{\uparrow}-n^{c}_{\downarrow} \right) \right>
%\end{eqnarray}
%%%%%%%%%%%%%%%%%%%%%%%%%%%%%%%%%%%%%%%%%%%%%%%%%%%%%%%%%%%%%%%%%%%%%%%%%%%%%%%
Introducing $U_{c}$,  the double occupation probabilities 
of the two bands are decreased, and
accordingly the {\it c-f} spin
 correlation function approaches -3/4, implying that
the spin singlet is formed between 
the $f$ electron and the conduction electron.
Therefore, the spin sector 
for large $U_f$ and  $U_c$ is stabilized by the formation 
of the local singlet at each site.

\begin{figure}[htb]
\begin{center}
\vskip 3mm
\leavevmode \epsfxsize=7cm 
\epsffile{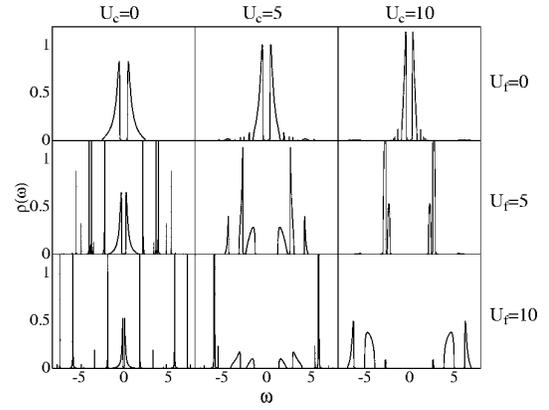}
\end{center}
\caption{ 
 DOS of $f$ electrons for the single-band PAM.
%The data are obtained by the linearized DMFT.
%%% with $\delta = 0.002$. 
}
\label{fig:single-dos}
\end{figure}

Although the interaction $U_c$ enhances electron correlations,
 it may not necessarily reduce 
the spectral gap characteristic 
of the Kondo insulating phase.  To see this clearly, we compute 
the density of states (DOS) for the 
one-particle spectrum  shown in Fig. \ref{fig:single-dos}.
%%%%%%%%%%%%%%%%%%%%%%%%%%%%%%%%%%%%%%%%%%%%%%%%%%%%%%%%%%%%%%%%%%%%%%%%
%%%%%%%%%%%%%%%%%%%%%%%%%%%%%%%%%%%%%%%%%%%%%%%%%%%%%%%%%%%%%%%%%%%%%%%%%%%%%%
For the non-interacting case ($U_{f}=U_{c}=0$), the gap 
in the vicinity of the Fermi level is 
caused by the bare {\it c-f} hybridization. 
When the {\it f-f} Coulomb interaction $U_{f}$ is increased
by keeping $U_c=0$, 
$f$ electrons are renormalized, resulting in
the formation of the small gap typical for the Kondo insulator. 
In addition to the low-energy Kondo peaks,
the large spectral weight appears around
$\tilde{\varepsilon}_{f}$ and $\tilde{\varepsilon}_{f}+U_f$,
which is represented by a bunch of spiky peaks
since the effective heat bath is represented by a few sites
in the linearized DMFT.
Similar behavior to reduce the gap is observed when only the {\it c-c} 
Coulomb interaction $U_c$ is introduced, as should be expected.

\begin{figure}[htb]
\begin{center}
\vskip 5mm
\leavevmode \epsfxsize=65mm 
\epsffile{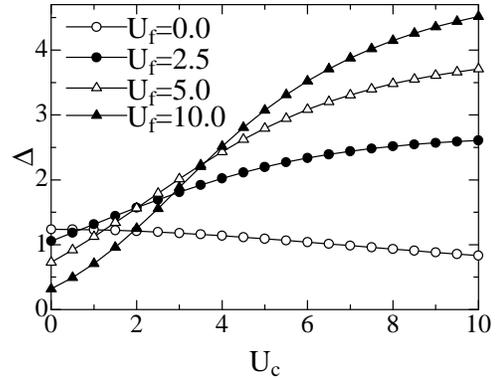}
\end{center}
\caption{
Spectral gap $\Delta$ for different values of $U_{f}$ as a
 function of 
the {\it c-c} interaction $U_{c}$ among conduction electrons.
}
\label{fig:single-gap}
\end{figure}

On the other hand, when  $U_c$ is turned on together with $U_f$,
the gap is {\it increased} as shown in Fig. \ref{fig:single-dos}.
This is contrasted to the above-mentioned cases possessing either 
of $U_f$ or $U_c$. To make this behavior more explicit, we plot 
the gap $\Delta$ in Fig. \ref{fig:single-gap}, from 
which we can indeed see the above mentioned properties.
%%%%%%%%%%%%%%%%%%%%%%%%%%%%%%%%%%%%%%%%%%%%%%%%%%%%%%%%%%%%%%%%%
%%%%%%%%%%%%%%%%%%%%%%%%%%%%%%%%%%%%%%%%%%%%%%%%%%%%%%%%%%%%%%%%%%%
The gap is expressed in terms of the renormalization factors as
%%%%%%%%%%%%%%%%%%%%%%%%%%%%%%%%%%%%%%%%%%%%%%%%%%%%%%%%%%%%%%%%%%
\begin{eqnarray}
\Delta=-Z_{c}+\sqrt{{Z_{c}}^2+4Z_{c}Z_{f}V_{g}^{2}},
\label{hybri}
\end{eqnarray}
%%%%%%%%%%%%%%%%%%%%%%%%%%%%%%%%%%%%%%%%%%%%%%%%%%%%%%%%%%%%%%%%%
where $V_g \equiv V+g$ is the effective hybridization
defined before.
It is seen from this formula that both of $Z_f$ and $Z_c$
have a tendency to reduce the gap $\Delta$. For example, 
when $U_c=0$, the gap 
 is decreased by $U_f$, leading to the Kondo insulator with
a small gap  $\sim 4Z_f V^2$. 
As mentioned above, however, the interplay of $U_c$ and
$U_f$  increases the gap.  We find that this enlargement
is due to the {\it c-f}  non-diagonal  self-energy.
Namely, in contrast to the simple PAM, 
the hybridization gap  is affected  not 
only by  $Z_{f}$ and $Z_{c}$ but also by the 
shift in the {\it c-f} hybridization, $g$, via the 
relation $V_g=V+g$ in the formula (\ref{hybri}).
Note that $g$ is finite only 
in the case $U_c\neq 0$ and $U_f\neq 0$, 
and gives rise to 
the large gap in Fig. \ref{fig:single-gap}.
In this way, the Kondo insulator is adiabatically connected 
to the Mott insulator realized in the case $U_c, U_f >> t$,
 where the gap is mainly determined by  $U_f$  or $U_c$. 
These results are consistent with those obtained by 
non-crossing approximation.\cite{T.Schork}

%%%%%%%%%%%%%%%%%%%%%%%%%%%%%%%%%%%%%%%%%%%%%%%%%%%%%%%%%%%%%%%%%%%
\section{Two-Orbital Periodic Anderson Model}
%%%%%%%%%%%%%%%%%%%%%%%%%%%%%%%%%%%%%%%%%%%%%%%%%%%%%%%%%%%%%%%%%%%

We now move to the two-orbital PAM  to 
clarify the role of the  orbital degrees of freedom.

%%%%%%%%%%%%%%%%%%%%%%%%%%%%%%%%%%%%%%%%%%%%%%%%%%%%%%
\subsection{Correlations due to {\it f-f} interactions}
%%%%%%%%%%%%%%%%%%%%%%%%%%%%%%%%%%%%%%%%%%%%%%%%%%%%%%%

\begin{figure}[htb]
\vskip 5mm
\begin{center}
\leavevmode \epsfxsize=80mm 
\epsffile{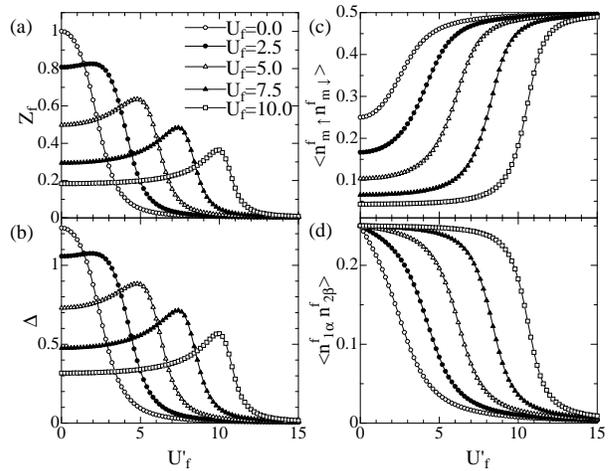}
\end{center}
\caption{
(a) Renormalization factor $Z_f$, (b) spectral gap $\Delta$, 
%(a) Spectral gap $\Delta$, (b)  renormalization factor $Z_f$, 
(c) and (d) local correlation functions for $f$ electrons, 
as a function of the inter-orbital {\it f-f} interaction $U'_{f}$
}
\label{fig:two-upf}
\end{figure}

We first turn off the
interactions for conduction electrons,
$U_{c}=U^{\prime}_{c}=J_c=0$, and  explore how the interactions for
$f$ electrons affect the formation of the Kondo insulator.
%%%%%%%%%%%%%%%%%%%%%%%%%%%%%%%%%%%%%%%%%%%%%%%%%%%%%%%%%%%%%%%%%%%
\begin{figure}[htb]
\vskip 5mm
\begin{center}
\leavevmode \epsfxsize=65mm 
\epsffile{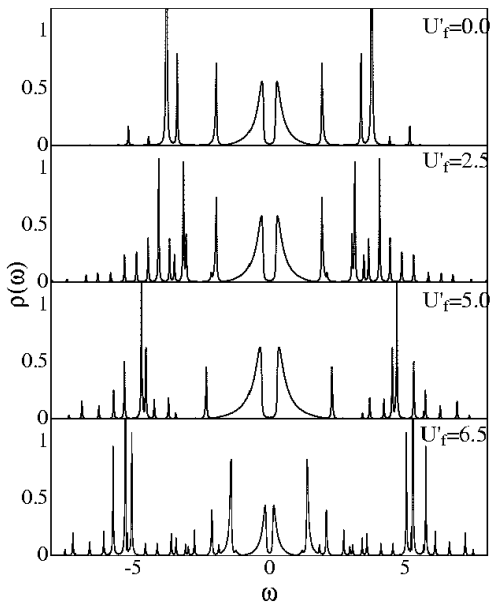}
\end{center}
\caption{
DOS of $f$ electrons 
for different values of the inter-orbital {\it f-f} Coulomb 
repulsion $U_{f}^{\prime}$
with a fixed $U_{f}=5.0$.
%These obtained by linearized DMFT with $\delta=0.015$. 
}
\label{fig:two-dos}
\end{figure}
%%%%%%%%%%%%%%%%%%%%%%%%%%%%%%%%%%%%%%%%%%%%%%%%%%%%%%%%%%%%%%%%%%%
Shown in Fig. \ref{fig:two-dos} is the DOS for $f$ electrons
when the inter-orbital {\it f-f} interaction $U'_f$ is varied
with $U_f=5.0$ being fixed.
Increasing $U_{f}^{\prime}$, the upper and lower Hubbard-type bands, 
which are represented by $\delta$-function like peaks,
are shifted to the higher energy region with their
weight being increased.  On the other hand, 
 qualitatively different behavior emerges
for the gap formation in the lower energy region.
Namely,  the size of the gap is once enhanced with 
the increase of $U_f^{\prime}$, but 
beyond $U_{f}^{\prime} \sim U_{f}$
it starts to decrease. This nonmonotonic behavior is explicitly
observed  in Figs. \ref{fig:two-upf} (a) and (b), where the 
renormalization factor $Z_f$ and the  size of the gap $\Delta$
are plotted as a function of  $U_f^{\prime}$.
%%%%%%%%%%%%%%%%%%%%%%%%%%%%%%%%%%%%%%%%%%%%%%%%%%%%%%%%%%%%%%%%%%%
%%%%%%%%%%%%%%%%%%%%%%%%%%%%%%%%%%%%%%%%%%%%%%%%%%%%%%%%%%%%%%%%%%%
In the case $U_f=0$, $Z_f$ and $\Delta$ decrease monotonically with 
the increase of $U'_f$. The resulting insulating phase
is regarded as a variant of the Kondo insulator, for which 
enhanced orbital (instead of spin) fluctuations reduce the gap.
On the other hand,  as already mentioned, a
finite $U_f$ leads to the maximum structure
in $Z_f$ and $\Delta$  around  $U'_f\sim U_f$.  
Therefore, low-energy electronic properties 
are quite sensitive to the balance of the Coulomb interactions 
$U_f$ and $U'_f$, as pointed out in a related 
context of the Hubbard model.\cite{A.Koga1,A.Koga2}
To confirm this, we also calculate the local correlation functions 
shown in Figs. \ref{fig:two-upf} (c) and (d).
The inter-orbital {\it f-f} interaction $U'_f$ increases
the double-occupation probability 
$\langle n^f_{m\uparrow}n^f_{m\downarrow} \rangle$ in the same orbital , 
while it decreases  $\langle n^f_{1\alpha}n^f_{2\beta} \rangle$
 for two electrons occupying different orbitals.  
It should be noticed  that these quantities are altered dramatically
around $U_f\sim U'_f$, implying that orbital fluctuations are 
enhanced there. Therefore, we can say that 
 the Kondo-insulating 
gap is enlarged due to orbital fluctuations among $f$ electrons.

We next consider the effects of the exchange coupling $J_{f}$ to 
further clarify 
the role of local spin and orbital degrees of freedom. 
%%%%%%%%%%%%%%%%%%%%%%%%%%%%%%%%%%%%%%%%%%%%%%%%%%%%%%%%%%%%%%%%%%%
\begin{figure}[htb]
\vskip 5mm
\begin{center}\leavevmode \epsfxsize=70mm 
\epsffile{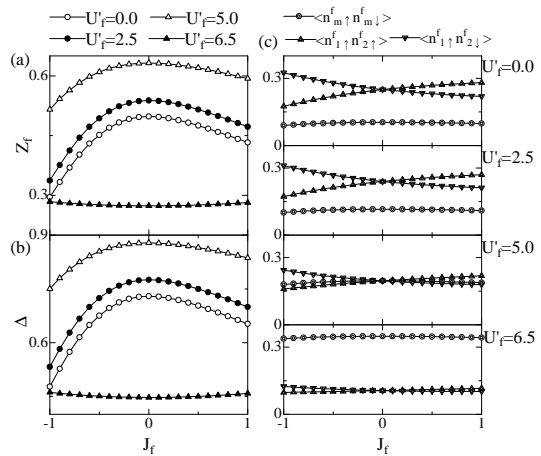}
\end{center}
\caption{
(a) Renormalization factor $Z_f$, (b) spectral gap $\Delta$, 
%(a) Spectral gap $\Delta$, (b) renormalization 
 and (c) local correlation 
functions as a function of the Hund coupling $J_{f}$. 
Other parameters are chosen as
%$U_{f}=U^{\prime}_{f}=10.0, U_{c}=U^{\prime}_{c}=0.0$ 
$U_{f}=10.0, U_{c}=U^{\prime}_{c}=0.0$ 
and $J_{c}=0.0$. 
}
\label{fig:two-jhf}
\end{figure}
%%%%%%%%%%%%%%%%%%%%%%%%%%%%%%%%%%%%%%%%%%%%%%%%%%%%%%%%%%%%%%%%%%%
In Figs. \ref{fig:two-jhf} (a) and (b), we show $Z_f$ and $\Delta$,
which exhibit analogous $J_f$ dependence.  Since
the Hund coupling restricts the available phase space
of $f$ electrons at each site, it has a tendency to suppress
 orbital fluctuations, thereby
reducing both of  $Z_f$ and $\Delta$.
Since the effective internal degrees of freedom 
depend on the sign of $J_f$,
the above effects appear differently
in two cases.
  When $J_f >0$, the Hund coupling favors a triplet state at each site,  
while the negative coupling $J_f(<0)$ a singlet state. 
Therefore, for $J_f <0$, orbital
fluctuations are suppressed rather strongly, thus 
leading to a more prominent decrease in $Z_f$ and $\Delta$.
If $U'_f$ gets large, the system favors the local configuration
of two electrons occupying the same orbital, so that 
the gap becomes insensitive to the Hund coupling 
in that parameter region, as seen in Figs. \ref{fig:two-jhf} (a) and (b).

For reference, we show  the local correlation functions 
$\langle n^{f}_{1\uparrow}n^{f}_{2\uparrow} \rangle$ and 
$\langle n^{f}_{1\uparrow}n^{f}_{2\downarrow} \rangle$
 in Fig. \ref{fig:two-jhf} (c),
from which we can check to what extent the exchange coupling $J_f$ affects
spin and orbital fluctuations.
It is seen that these quantities characterize the 
enhancement of triplet or singlet correlations
depending on the sign of $J_f$, but only 
in the region $U'_f<U_f$.
In the case $U^{\prime}_{f}> U_{f}$, however, the double occupancy 
of either of two orbitals is favored, and thus
$\langle n^{f}_{m\uparrow}n^{f}_{m\downarrow} \rangle$ gets larger than
$\langle n^{f}_{1\uparrow}n^{f}_{2\uparrow} \rangle$ and 
$\langle n^{f}_{1\uparrow}n^{f}_{2\downarrow} \rangle$.
This implies that the Hund coupling $J_{f}$ is irrelevant in this case,
consistent with the above results.

%%%%%%%%%%%%%%%%%%%%%%%%%%%%%%%%%%%%%%%%%%%%%%%%%%%%%%%%%%
\subsection{Correlations due to {\it c-c} interactions}
%%%%%%%%%%%%%%%%%%%%%%%%%%%%%%%%%%%%%%%%%%%%%%%%%%%%%%%%%%

%%%%%%%%%%%%%%%%%%%%%%%%%%%%%%%%%%%%%%%%%%%%%%%%%%%%%%%%%%%%%%%%%%%
\begin{figure}[htb]
\vskip 5mm
\begin{center}
\leavevmode \epsfxsize=70mm 
\epsffile{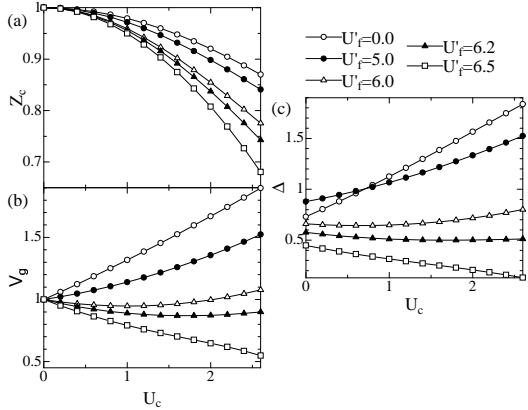}
\end{center}
\caption{
(a) Renormalization factor $Z_c$, (b) effective hybridization $V_g$ and 
(c) spectral gap $\Delta$
as a function of the intra-orbital {\it c-c} interaction $U_{c}$ 
for different values of the inter-orbital 
{\it f-f} interaction  $U^{\prime}_f$.
The intra-orbital {\it f-f} Coulomb interaction is fixed as 
$U_{f}=5.0$. 
}
\label{fig:two-uc-abc}
\end{figure}
%%%%%%%%%%%%%%%%%%%%%%%%%%%%%%%%%%%%%%%%%%%%%%%%%%%%%%%%%%%%%%%%%%%

%%%%%%%%%%%%%%%%%%%%%%%%%%%%%%%%%%%%%%%%%%%%%%%%%%%%%%%%%%%%%%%%%%%
\begin{figure}[htb]
\vskip 5mm
\begin{center}
\leavevmode \epsfxsize=70mm 
\epsffile{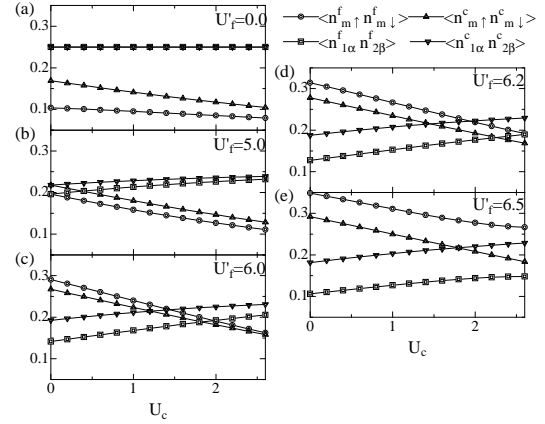}
\end{center}
\caption{
Local correlation functions  
as a function of intra-orbital {\it c-c} interaction $U_{c}$. 
The parameters are chosen as in \ref{fig:two-uc-abc}
}
\label{fig:two-uc-d}
\end{figure}
%%%%%%%%%%%%%%%%%%%%%%%%%%%%%%%%%%%%%%%%%%%%%%%%%%%%%%%%%%%%%%

We now study the influence coming from 
 correlations among conduction electrons. 
In Figs. \ref{fig:two-uc-abc} and \ref{fig:two-uc-d}, several quantities 
discussed so far are plotted in the presence 
of the {\it c-c} Coulomb interaction  $U_{c}$
 together with  
{\it f-f} interactions $U_{f}$ and $U^{\prime}_{f}$. 
We notice that the effect of $U_c$ shows up differently
in  the cases of $U_{f} > U^{\prime}_{f}$
and $U_{f} < U^{\prime}_{f}$. 
For $U_{f} > U^{\prime}_{f}$, 
the gap $\Delta$ increases, although
the renormalization factor $Z_c$
 decreases, reflecting electron correlations among
conduction electrons
(Figs. \ref{fig:two-uc-abc} (a) and (c)). This behavior is somewhat 
similar to the case 
 discussed in the single-orbital model in 
Figs. \ref{fig:single-renor} and \ref{fig:single-dos}.  
Namely, the increase of the gap is caused by 
 the increase of the hybridization $V_g=V+g$  due to
 the {\it c-f} self-energy. This  is indeed confirmed 
 in Fig. \ref{fig:two-uc-abc} (b). In this case, the system
changes continuously from the Kondo insulator 
 to the Mott insulator This
 is typically seen in the local correlation functions shown 
in Fig. \ref{fig:two-uc-d} (a), where both of the double-occupation
probabilities 
 $\langle n^{f}_{m\uparrow}n^{f}_{m\downarrow} \rangle$
and $\langle n^{c}_{m\uparrow}n^{c}_{m\downarrow} \rangle$
are suppressed with the increase of $U_c$.

On the other hand, in the case of $U_{f} < U^{\prime}_{f}$, 
the gap $\Delta$ 
 decreases  with the increase of $U_c$,
 which is caused by the decrease of $V_g$.  
As shown in  Figs. \ref{fig:two-uc-d} (d) and (e), 
the double-occupation probability 
of the same orbital at $U_c=0$ is enhanced not only for $f$ electrons
($\langle n^{f}_{m\uparrow}n^{f}_{m\downarrow} \rangle$)
but also for conduction electrons 
($\langle n^{c}_{m\uparrow}n^{c}_{m\downarrow} \rangle$),
even if $U_c'=0$.
This is a nontrivial effect due to the interplay of 
these interactions, and implies that the system at $U_c=0$
is a variant of the Kondo insulator, for which the strong
renormalization is caused by orbital fluctuations
due to the inter-orbital interactions.
The increase of $U_c$ in turn suppresses such enhanced orbital
fluctuations, as typically seen in Figs. \ref{fig:two-uc-d} (d) and (e),
and at the same time gives rise to 
the decrease of $V_g$.  Therefore, we may say that 
 the resulting insulator is still in the Kondo insulating
phase with a small gap, which possesses both of the 
enhanced  spin and orbital fluctuations.

%%%%%%%%%%%%%%%%%%%%%%%%%%%%%%%%%%%%%%%%%%%%%%%%%%%%%%%%%%%%%%%%%%%
%\vspace{0.5cm}
\begin{figure}[htb]
\vskip 5mm
\begin{center}
%\leavevmode \epsfxsize=35mm 
%\epsffile{two-upc.eps}
\leavevmode \epsfxsize=70mm 
\epsffile{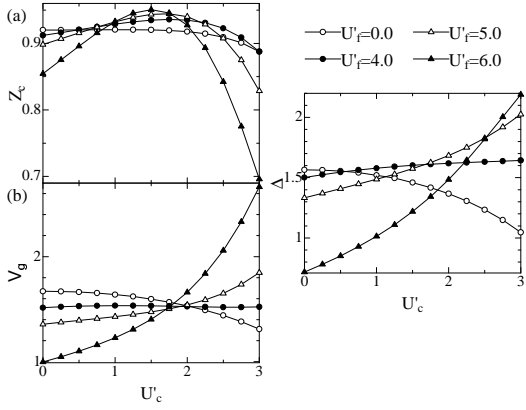}
\end{center}
\caption{
(a) Renormalization factor of conduction electrons $Z_c$, 
(b) effective hybridization $V_g$, (c)
spectral gap $\Delta$ 
as a function of the inter-orbital {\it c-c} Coulomb interaction 
$U^{\prime}_{c}$.  
We set $U_{f}=5.0$ and $U_{c}=2.0$ for simplicity.
}
\label{fig:two-upc-abc}
\end{figure}
%%%%%%%%%%%%%%%%%%%%%%%%%%%%%%%%%%%%%%%%%%%%%%%%%%%%%%%%%%%%%%%%%%%

%%%%%%%%%%%%%%%%%%%%%%%%%%%%%%%%%%%%%%%%%%%%%%%%%%%%%%%%%%%%%%%%%%%
%\vspace{0.5cm}
\begin{figure}[htb]
\vskip 5mm
\begin{center}
%\leavevmode \epsfxsize=35mm 
%\epsffile{two-upc.eps}
\leavevmode \epsfxsize=70mm 
\epsffile{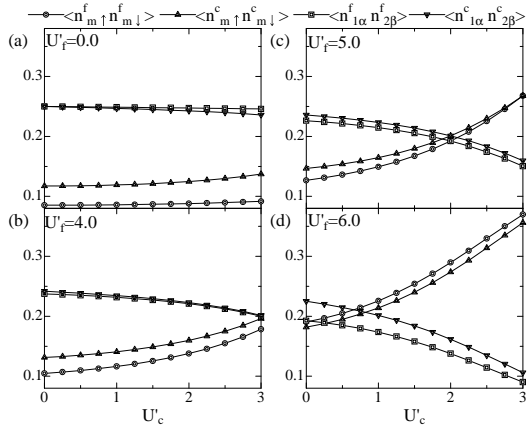}
\end{center}
\caption{
Local correlation functions  for
$f$ and conduction electrons.
%% on the single orbital 
%%$\langle n^{f}_{m\uparrow}n^{f}_{m\downarrow} \rangle$ 
%%($\langle n^{c}_{m\uparrow}n^{c}_{m\downarrow} \rangle$) 
%%and the double occupancy of $f$ (conduction) electron 
%%on the different orbital $\langle n^{f}_{1\sigma} 
%%n^{f}_{2\sigma^{\prime}} \rangle$ 
%%($\langle n^{c}_{1\sigma} n^{c}_{2\sigma^{\prime}} \rangle$).
We set $U_{f}=5.0$ and $U_{c}=2.0$. 
}
\label{fig:two-upc-d}
\end{figure}
%%%%%%%%%%%%%%%%%%%%%%%%%%%%%%%%%%%%%%%%%%%%%%%%%%%%%%%%

We finally observe what happens when we further
introduce the inter-orbital {\it c-c} 
Coulomb interaction $U_c'$ among conduction electrons. 
We fix the values $U_{f}=5.0$ and $U_{c}=2.0$, and 
compute the gap and the renormalization factor of 
conduction electrons as a function of $U^{\prime}_{c}$ for various 
choices  of $U^{\prime}_{f}$.  Since we can
follow the arguments outlined above, we only mention
some interesting points briefly. As seen from
 Fig. \ref{fig:two-upc-abc}, the gap is not so sensitive to 
the value of $U_c'$ for small $U_f'$.
A remarkable point is that when $U_f'$ is large (e.g.
$U_f'=6.0$), all of $Z_c$, $V_g$ and $\Delta$ change 
dramatically as a function of $U_c'$.
Although the spectral gap $\Delta$ increases 
similarly to the case discussed in Fig. \ref{fig:two-uc-abc}
for small $U_f'$,
 we should notice that the nature of the insulating 
phase is  different from each other.
The difference is  distinguished in terms of
the local correlation functions
shown in Fig. \ref{fig:two-upc-d}.   In particular, from panel (d)
for $U_f'=6.0$, we can see that 
$\langle n^{f}_{m\uparrow}n^{f}_{m\downarrow} \rangle$
and $\langle n^{c}_{m\uparrow}n^{c}_{m\downarrow} \rangle$ 
are enhanced,  whereas 
$\langle n^{f}_{1\alpha} n^{f}_{2\beta} \rangle$ and
 $\langle n^{c}_{1\alpha} n^{c}_{2\beta} \rangle$ are suppressed. 
This implies that the insulating phase in that parameter region 
 may be regarded as a variant of
the Mott insulator with a large gap, for which 
orbital fluctuations are enhanced while spin fluctuations
are suppressed.  This phase is qualitatively different from the 
one discussed in Fig. \ref{fig:two-uc-abc}
for small $U_f'$, where spin (orbital) fluctuations are
enhanced (suppressed) in the presence of $U_c$.

Before closing this section, brief comments are in order for the impacts
due to other interactions we have not addressed here.
We have refrained from discussing 
the effects due to the exchange coupling $J_c$ among conduction 
electrons, 
since they are mostly the same as those of $J_f$.  
Also, we have not argued the influence due to the {\it c-f}
Coulomb repulsion in order to avoid the model to be too complicated.  
We have checked that 
this interaction has a tendency to enlarge the spectral gap.

%%%%%%%%%%%%%%%%%%%%%%%%%%%%%%%%%%%%%%%%%%%%%%%%%%%%%%%%%%%%%%%%%%%%%%%%%%
%%%%%%%%%%%%%%%%%%%                          %%%%%%%%%%%%%%%%%%%%%%%%%%%%%
%%%%%%%%%%%%%%%%%%%    4. conclusion         %%%%%%%%%%%%%%%%%%%%%%%%%%%%%
%%%%%%%%%%%%%%%%%%%                          %%%%%%%%%%%%%%%%%%%%%%%%%%%%%
%%%%%%%%%%%%%%%%%%%%%%%%%%%%%%%%%%%%%%%%%%%%%%%%%%%%%%%%%%%%%%%%%%%%%%%%%%

\section{Summary}

We have studied electron correlations for the orbitally 
degenerate periodic Anderson model 
by means of a linearized version of dynamical mean field theory.
In particular, we have focused on the role played by  
 various intra- and inter-orbital interactions. 

By taking the single orbital model, we have first demonstrated how  
 the {\it c-c} Coulomb repulsion  naturally interpolates the
Kondo insulator and the Mott insulator. 
In this case, the {\it c-f} part of the self-energy plays
a major role rather than the renormalization effects due to
$Z_c$ and $Z_f$. 

In the two orbital model,  there are some remarkable 
effects, which do not appear in the single-orbital case,
due to the interplay of the intra- and inter-orbital
interactions.  One of the most interesting 
results is that orbital fluctuations are enhanced
around $U_{f}=U^{\prime}_{f}$, which in turn 
 suppress the renormalization effect. As a result,
 the Kondo insulating gap shows a maximum around there.
This effect is somehow obscured by
the exchange coupling $J_{f}$ between $f$ orbitals
 since it has a tendency to reduce 
orbital fluctuations. Accordingly, the introduction of 
$J_f$  again enhances the renormalization
effect, making the Kondo gap smaller.

Upon introducing the interactions for conduction electrons as well as 
$f$ electrons in the degenerate model, there appear a 
variety of  remarkable effects due to
the interplay of these interactions. In fact, reflecting 
the subtle balance of the interactions, the insulating phase with 
strong correlations has either the enlarged gap or the reduced gap, 
for which  either of spin or orbital fluctuations 
(both of them in some cases) are enhanced.

%%Although we have used a simplified version of the DMFT in this
%%paper, we think that
%%the present analysis should capture the essential properties due to
%% the interplay of various interactions in the orbitally degenerate
%%PAM.  

In this paper, we have restricted our analysis to the non-magnetic phase.
We have indeed observed  a number of 
interesting  properties even in such nonmagnetic insulating phase.
It remains an interesting problem in the future study to clarify the 
competition between the nonmagnetic insulating state and magnetically
ordered states, which may be particularly important in the 
system possessing large interactions.

%%%%%%%%%%%%%%%%%%%%%%%%%%%%%%%%%%
\section*{Acknowledgements}
%%%%%%%%%%%%%%%%%%%%%%%%%%%%%%%%%%
The authors thank T. Mutou for valuable discussions.
This work was partly supported by a Grant-in-Aid from the Ministry 
of Education, Science, Sports and Culture of Japan. 
A part of computations was done at the Supercomputer Center at the 
Institute for Solid State Physics, University of Tokyo
and Yukawa Institute Computer Facility.

%%%%%%%%%%%%%%%%%%%%%%%%%%%%%%%%%%%%%%%%%%%%%%%%%%%%%%%%%%%%%%%%%%%
%%%%%%%%%                             %%%%%%%%%%%%%%%%%%%%%%%%%%%%%
%%%%%%%%%      Reference              %%%%%%%%%%%%%%%%%%%%%%%%%%%%%
%%%%%%%%%                             %%%%%%%%%%%%%%%%%%%%%%%%%%%%%
%%%%%%%%%%%%%%%%%%%%%%%%%%%%%%%%%%%%%%%%%%%%%%%%%%%%%%%%%%%%%%%%%%%


\begin{thebibliography}{99}

%------------------------------------
% Kondo Insulator (YbB12)
%------------------------------------
\bibitem{T.Susaki}
T. Susaki {\it et al}: Phys. Rev. Lett. {\bf 77} (1996) 4269.
%
%
\bibitem{K.Sugiyama}
K. Sugiyama, F Iga, and M. Kasaya: J. Phys. Soc. Jpn. {\bf 57} (1988) 3946. 
%
%
\bibitem{T.Saso-1}
T. Saso and H. Harima: J. Phys. Soc. Jpn. {\bf 68} (1999) 2491. 
%
%
%------------------------------------
% Kondo Insulator
%------------------------------------

\bibitem{J.W.Allen}
J. W. Allen, B. Batlogg, and P. Wachter: Phys. Rev. B. {\bf 20} (1979) 4807. 

\bibitem{M.F.Hundley}
M. F. Hundley, P. C. Canfield, and Z. Fisk: Phys. Rev. B. {\bf 42} (1990) 6842. 

\bibitem{A.Severing}
A. Severing, J. D. Thompson, P. C. Canfield, Z. Fisk, and P. S. Riseborough: 
Phys. Rev. B {\bf 44} (1991) 6832.

\bibitem{T.E.Mason}
T. E. Mason {\it et al}: Phys. Rev. Lett. {\bf 69} (1992) 490. 

\bibitem{P.S.Riseborough}
P. S. Riseborough: Phys. Rev. B {\bf 45} (1992) 13984.

\bibitem{C.S.Castro}
C. Sanchez-Castro, K. S. Bedell, and B. R. Cooper: 
Phys. Rev. B {\bf 47} (1993) 6879.

\bibitem{J.C.Cooley}
J. C. Cooley, M. C. Aronson, Z. Fisk, and P. C. Canfield: 
Phys. Rev. Lett. {\bf 74} (1995) 1629.

%------------------------------------
% Nd
%------------------------------------
\bibitem{T.Brugger}
T. Brugger, T. Schreiner, G. Roth, P. Adelmann, and G. Czjzek: 
Phys. Rev. Lett. {\bf 71} (1993) 2481.

\bibitem{Y.M.Li}
Y. M. Li: Phys. Rev. B. {\bf 52} (1995) 6979.

\bibitem{J.Igarashi}
J. Igarashi, K. Murayama, and P. Fulde: Phys. Rev. B. {\bf 52} (1995) 15966.

\bibitem{K.Itai}
K. Itai and P. Fazekas: Phys. Rev. B. {\bf 54} (1996) 752.

\bibitem{T.Schork}
T. Schork and S. Blawid: Phys. Rev. B. {\bf 56} (1997) 6559.
%
%
%

%------------------------------------
% multichannel
%------------------------------------
\bibitem{Y.Ono-1}
Y. Ono, T. Matsuura, and Y. Kuroda:J. Phys. Soc. Jpn. {\bf 63} (1993) 1406.

\bibitem{M.Jarrell-1}
M. Jarrell, H. Pang, D. L. Cox, F. Anders, and A. Chattopadhyay: 
Physica B {\bf 230-232} (1997) 557.

\bibitem{A.Tsuruta}
A. Tsuruta, A. Kobayashi, and Y. \=Ono: J. Phys. Soc. Jpn. {\bf 68} (1999) 2491.

%------------------------------------
% DMFT 
%------------------------------------

\bibitem{W.Metzner}
W. Metzner and D. Vollhardt: Phys. Rev. Lett. {\bf 69} (1989) 324.

\bibitem{MullerHartmann}
E. M\"uller-Hartmann, Z. Phys. B: Condens. Matter {\bf 74} (1989) 507.

\bibitem{Pruschke}
T. Pruschke, M. Jarrell, and J. K. Freericks: Adv. Phy. {\bf 42} (1995) 187.

\bibitem{Rev}
A. Georges, G. Kotliar, W. Krauth, and M. J. Rozenberg: 
Rev. Mod. Phys {\bf 68} (1996) 13.


%
%
%------------------------------------
% DMFT (single-Hubbard)
%------------------------------------

\bibitem{A.Georges}
A. Georges and G. Kotliar: Phys. Rev. B. {\bf 45} (1992) 6479.

\bibitem{M.Jarrell-2}
M. Jarrell: Phys. Rev. Lett. {\bf 69} (1992) 168.

\bibitem{X.Y.Zhang}
X. Y. Zhang, M. J. Rozenberg, and G. Kotliar: 
Phys. Rev. Lett. {\bf 70} (1993) 1666.

\bibitem{M.Caffarel}
M. Caffarel and W. Krauth: Phys. Rev. Lett. {\bf 72} (1994) 1545.

\bibitem{D.S.Fisher}
D. S. Fisher, G. Kotliar, and G. Moeller: 
Phys. Rev. B. {\bf 52} (1995) 17112.

\bibitem{H.Kajueter}
H. Kajueter and G. Kotliar: Phys. Rev. Lett. {\bf 77} (1996) 131.

\bibitem{R.Bulla-1}
R. Bulla: Phys. Rev. Lett. {\bf 83} (1999) 136.

\bibitem{M.Potthoff-1}
M. Potthoff: Phys. Rev. B. {\bf 64} (2000) 165114.

\bibitem{R.Bulla-2}
R. Bulla, and M. Potthoff: Eur. Phys. J. B. {\bf 13} (2000) 257.

\bibitem{Y.Ono-2}
Y. \=Ono, R. Bulla, A. C. Hewson, and M. Potthoff:
Eur. Phys. J. B. {\bf 22} (2001) 283.



%------------------------------------
% DMFT (two-Hubbard)
%------------------------------------
\bibitem{M.J.Rozenberg}
M. J. Rozenberg: Phys. Rev. B. {\bf 55} (1997) R4855.

\bibitem{K.Held}
K. Held and D. Vollhardt: Euro. Phys. J. B {\bf 5} (1998) 473.

\bibitem{J.E.Han}
J. E. Han, M. Jarrell, and D. L. Cox: Phys. Rev. B. {\bf 58} (1998) R4199.

\bibitem{W.Metzner2}
Th. Maier, M. B. Z\"olfl, Th. Pruschke, and J. Keller:
Eur. Phys. J. B {\bf 19} (1999) 377.
%
%
\bibitem{T.Momoi}
T. Momoi and K. Kubo: Phys. Rev. B. {\bf 58} (2000) R567.

\bibitem{Y.Imai}
Y. Imai and N. Kawakami: J. Phys. Soc. Jpn. {\bf 70} (2001) 2365.
%

\bibitem{A.Koga1}
A. Koga, Y. Imai, and N. Kawakami: Phys. Rev. B. {\bf 66} (2002) 165107.

\bibitem{V.S.Oudovenko}
V. S. Oudovenko and G. Kotliar: Phys. Rev. B. {\bf 65} (2002) 0750102.

\bibitem{S.Florens}
S. Florens, A. Georges, G. Kotliar, and O. Parcollet: 
Phys. Rev. B. {\bf 66} (2002) 205102.

\bibitem{Y.Ono}
Y. \=Ono, M. Potthoff, and R. Bulla: Phys. Rev. B. {\bf 67} (2003) 035119.

\bibitem{A.Koga2}
A. Koga, T. Ohashi, Y. Imai, S. Suga and N. Kawakami: J. Phys. Soc. Jpn.
{\bf 72} (2003) 1306.

\bibitem{Liebsch}
A. Liebsch: Phys. Rev. Lett. {\bf 91} (2003) 226401.

\bibitem{A.Koga3}
A. Koga, N. Kawakami, T.M. Rice, and M. Sigrist: cond-mat/0401223.


%
%------------------------------------
% DMFT (periodic Anderson)
%------------------------------------
\bibitem{T.Mutou}
T. Mutou and D. Hirashima: J. Phys. Soc. Jpn. {\bf 63} (1994) 4475;
T. Mutou: Phys. Rev. B. {\bf 64} (2001) 165103.

\bibitem{M.Jarrell-3}
M. Jarrell: Phys. Rev. B. {\bf 51} (1995) 7429.

\bibitem{T.Saso-2}
T. Saso and M. Itoh: Phys. Rev. B. {\bf 53} (1996) 6877.

\bibitem{Th.Pruschke}
Th. Pruschke, R. Bulla, M. Jarrell: Phys. Rev. B. {\bf 61} (2000) 12799.

\bibitem{D.Meyer}
D. Meyer and W. Nolting: Phys. Rev. B. {\bf 61} (2000) 13465.

%------------------------------------
% DMFT (NCA)
%------------------------------------
\bibitem{C.I.Kim}
Chang-Il Kim, Y. Kuramoto, and T. Kasuya: 
J. Phys. Soc. Jpn. {\bf 59} (1990) 2414.

%%%%%%%%%%%%%%%%%%%%%%%%%%%%

\end{thebibliography}
\end{document}